\documentclass{article}
\usepackage{spconf,amsmath,graphicx}
\usepackage{booktabs}
\usepackage{makecell}
\usepackage{multirow}
\usepackage{tabularx}
\usepackage{array}
\usepackage{xcolor}
\usepackage{makecell}
\usepackage{pifont}
\usepackage{url}
\usepackage{cite}
\usepackage{amssymb}

\title{CTC-Assisted LLM-Based Contextual ASR}
\name{
  Guanrou Yang$^{1}$,
  Ziyang Ma$^{1}$,
  Zhifu Gao$^{2}$,
  Shiliang Zhang$^{2}$,
  Xie Chen$^{1,\dagger}$
}

\address{
  $^1$MoE Key Lab of Artificial Intelligence, AI Institute, X-LANCE Lab, \\ Shanghai Jiao Tong University, China,
  $^2$Alibaba Group, China \\
\texttt{\{yangguanrou, chenxie95\}@sjtu.edu.cn}}

\newcommand\blfootnote[1]{
  \begingroup
  \renewcommand\thefootnote{}\footnote{#1}
  \addtocounter{footnote}{-1}
  \endgroup
}

\begin{document}
%
\maketitle
\begin{abstract}
\vspace{-1mm} 
Contextual ASR or hotword customization holds substantial practical value. Despite the impressive performance of current end-to-end (E2E) automatic speech recognition (ASR) systems, they often face challenges in accurately recognizing rare words. Typical E2E contextual ASR models commonly feature complex architectures and decoding mechanisms, limited in performance and susceptible to interference from distractor words. With large language model (LLM)-based ASR models emerging as the new mainstream, we propose a CTC-Assisted LLM-Based Contextual ASR model with an efficient filtering algorithm. By using coarse CTC decoding results to filter potential relevant hotwords and incorporating them into LLM prompt input, our model attains WER/B-WER of 1.27\%/3.67\% and 2.72\%/8.02\% on the Librispeech test-clean and test-other sets targeting on recognizing rare long-tail words, demonstrating significant improvements compared to the baseline LLM-based ASR model, and substantially surpassing other related work. More remarkably, with the help of the large language model and proposed filtering algorithm, our contextual ASR model still performs well with 2000 biasing words.\footnote{Code and checkpoints are available at 
 https://github.com/X-LANCE/SLAM-LLM/tree/main/examples/contextual\_asr}
\end{abstract}

\blfootnote{$\dagger$ Corresponding author}
\vspace{-4mm} 



%
\begin{keywords}
contextual speech recognition, large language model, contextual biasing
\end{keywords}

\begin{figure*}[t]
  \centering
  \includegraphics[width=\linewidth]{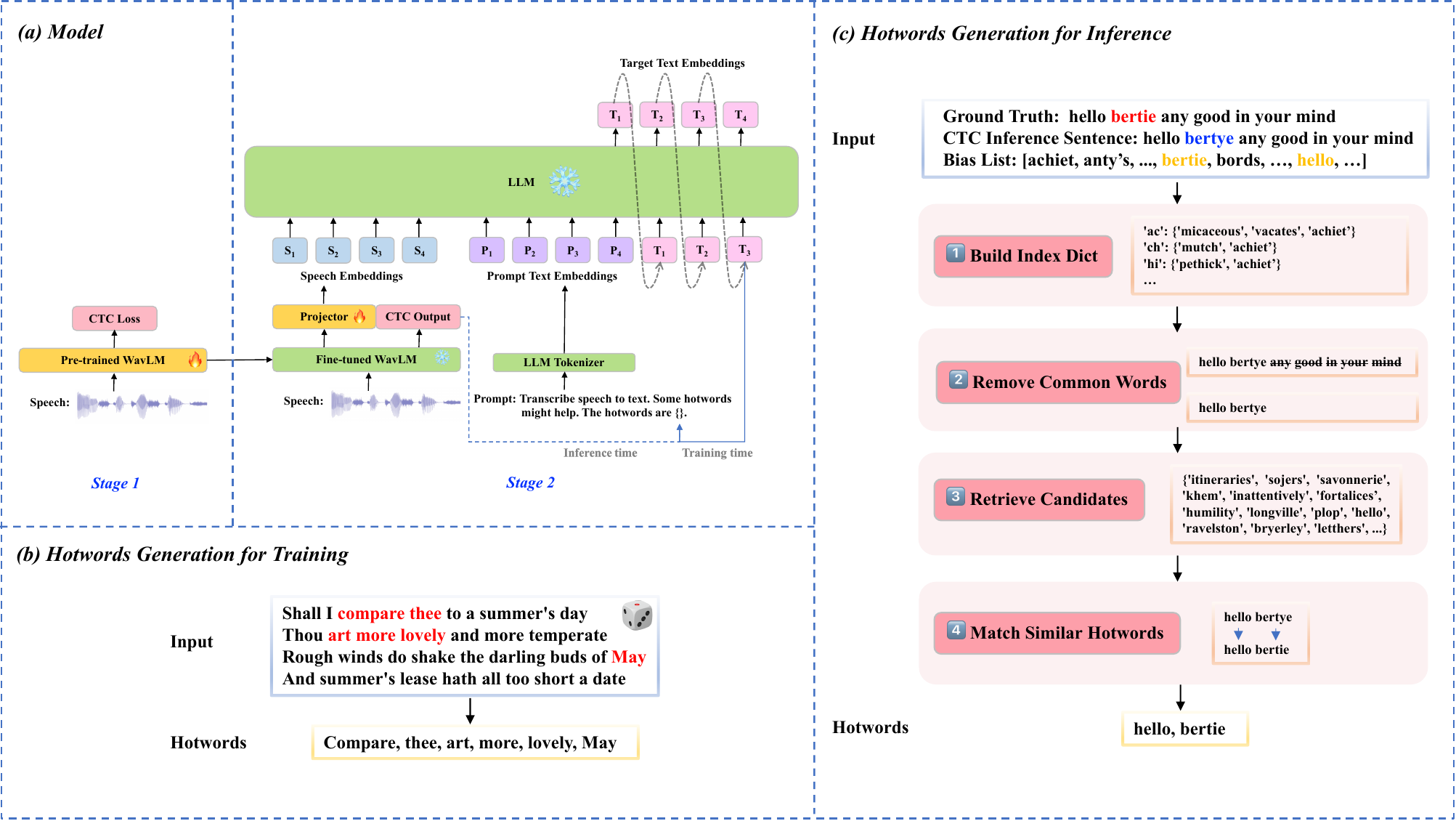}
  \caption{(a) illustrates the proposed Contextual LLM-based ASR model architecture. (b) and (c) explain the generation process of hotwords to be included in the prompt during training and inference phases respectively.}
  \label{fig:model}
  \vspace{-3mm}
\end{figure*}

\vspace{-4mm} 
\section{Introduction}
\vspace{-3mm} 
End-to-end (E2E) automatic speech recognition (ASR) systems demonstrate impressive performance~\cite{li2022recent} but often struggle with accurately recognizing rare phrases, such as technical terms and name entities that are infrequently encountered in training data. Integrating contextual knowledge into E2E ASR models for biased decoding is crucial for improving recognition accuracy in practical applications, such as commercial systems where users need personalized recognition for specific names, places, and entities, highlighting the academic and commercial value of supporting hotword customization.~\cite{shi2024seaco}

To address this challenge, researchers have developed various methods to inject contextual information into E2E models.
The traditional approaches involve shallow fusion~\cite{williams2018contextual,zhao2019shallow,chen2019end,huang2020class}, which utilizes an independently built language model (LM). 
Similarly, on-the-fly (OTF) rescoring method~\cite{pundak2018deep,hall2015composition} compiles known word-level biasing phrases into a weighted finite state transducer (WFST), combined with a ``speller" FST to convert grapheme or word-piece sequences into words. A contextual LM score derived from this combination is integrated into the decoding criterion of standard log-likelihood, controlled by a tunable hyperparameter.

In contrast, all-neural contextual biasing methods~\cite{pundak2018deep,jain2020contextual,huang2023contextualized,han2021cif,shi2024seaco} integrate an additional contextual module with the ASR module. For instance, CLAS~\cite{pundak2018deep} dynamically incorporates contextual information during speech recognition by embedding context phrases and integrating them into the ASR components via the attention mechanism. These methods can handle out-of-vocabulary terms effectively and do not demand careful tuning of hyperparameters like rescoring weights, leading to significant improvements over OTF rescoring.

In practical applications, a biasing list often includes hundreds to thousands of biasing words. Retrieving a meaningful bias context vector is challenging, for it's difficult for the cross-attention mechanism to accurately link the ASR decoder output with the large-scale sparse hotword embeddings. As the length of the biasing list increases and the number of interfering items grows, the model's performance tends to decline. To tackle this problem, different works have introduced various filtering methods for processing the biasing list tailored to their specific models. For instance, CLAS proposes a ``bias-conditioning" technique that activates a bias phrase only if its prefix matches part of the partially decoded hypothesis. A two-stage contextual word filter module~\cite{yang2023two} is introduced for attention-based context bias, which includes computing ``Posterior Sum Confidence" and ``Sequence Order Confidence", especially designed for cascaded encoder ASR framework.

The above approaches often involve intricate model structures and decoding processes. In this paper, inspired by the recent emergence of LLM-based ASR models~\cite{wu2023decoder,ma2024embarrassingly,yang2024mala, gong2024contextual} with simple structures and powerful performance, we propose an effective LLM-based contextual ASR model, along with a robust filtering algorithm.
Prior work MaLa-ASR~\cite{yang2024mala} has successfully demonstrated the ability of the LLM-based ASR model to integrate textual keywords extracted from presentation slides to improve recognition of conference content. 
Different from their AVSR task scenario, where each utterance corresponds to a slide with limited related content, contextual ASR only utilizes a single long biasing list typically containing thousands of entries in the inference stage. Thus, we propose a filtering algorithm where we leverage coarse decoding results generated by a fine-tuned SSL model with a simple CTC~\cite{graves2006connectionist} head to filter and select the most pertinent hotwords. Then the filtered hotwords are incorporated into the prompts input for LLM. Our proposed CTC-Assisted LLM-based contextual ASR model, trained and evaluated on Librispeech~\cite{panayotov2015librispeech} corpus, achieves WER/B-WER of 1.27\%/3.67\% on the test-clean set and 2.72\%/8.02\% on the test-other set with a biasing list size of 100, showing notable relative reductions of 39.81\%/63.37\% and 35.24\%/61.37\% compared to the baseline LLM-based ASR model, and significantly outperforming other related work.

\vspace{-4mm} 
\section{Method}
\vspace{-4mm} 
In this section, we introduce our model architecture and propose a filtering algorithm, concretely interpreted in Figure \ref{fig:model}.

\vspace{-3mm}
\subsection{Model Architecture}
\vspace{-2mm} 


Attaching a CTC head to a speech SSL pre-trained model allows for obtaining an initial decoding result with minimal additional computational overhead. Meanwhile, the fine-tuning process with CTC loss yields a better encoder for LLM-based ASR training, outperforming the pre-trained model. 
Thus, we propose an LLM-based speech model utilizing a CTC fine-tuned WavLM~\cite{chen2022wavlm} encoder for contextual ASR based on the framework of MaLa-ASR. 
The powerful SSL model WavLM Large is used for feature extraction, pre-trained on $94,000$ hours of data, and fine-tuned on $960h$ hours of Librispeech data. The Vicuna 7B~\cite{chiang2023vicuna} is used as the LLM decoder. A lightweight simple-structured linear projector is utilized for aligning the extracted speech feature with the LLM input space. It first downsamples the 50Hz features to 10Hz by a 1-D convolution layer and then projects it by two linear layers with an intermediate hidden layer dimension of 2048.

For contextual ASR, we incorporate a biasing list in the prompt to assist the LLM in accurately recognizing rare words that are often misidentified.
Given the differences in tasks and scenarios compared to previous LLM-based ASR models, we design a new task instruction and introduce novel methods for generating biasing lists.
The training biasing list is generated randomly from the transcriptions of speech in the current training batch, adhering to prior research~\cite{pundak2018deep,huang2023contextualized,futami2024phoneme}. Each transcription has a probability of $P_{keep}$ of being used to extract biasing words. For each kept transcription, $k$ word-level $n$-grams are randomly chosen, where $k$ is sampled uniformly from $[1, N_{phrases}]$ and $n$ is sampled uniformly from $[1, N_{order}]$, where $ N_{phrases}$ and $N_{order}$ are hyperparameters controling the number and length of n-grams.
The test biasing list consists of hotwords selected from a provided predefined biasing list containing hundreds to thousands of words, using the filtering algorithm described in the next subsection.

The overall training workflow is as follows: The WavLM encoder processes 16kHz input speech into 50Hz feature sequences of dimension 1024. These sequences are downsampled and projected into 10Hz speech embeddings with a dimension of 4096. The Prompt with task instruction and biasing list is tokenized and encoded into text embeddings. Both speech and prompt text embeddings are concatenated into a unified representation, fed into the LLM, and decoded to produce target transcription autoregressively. Only the lightweight projector is trained while the rest of the model remains frozen. Cross-entropy loss is calculated between hypothesis and transcription, following standard practice for training LLM-based speech models. 

\begin{table*}[ht]
\centering
\caption{ Performance of CTC-Assisted LLM-Based Contextual ASR on LibriSpeech test sets. ``No bias" refers to adding empty string, ``Bias List" means incorporating filtered hotwords from the complete biasing list, and ``GT Hotwords" means including the precise ground truth hotwords during inference.}
  \resizebox{\linewidth}{!}{
\begin{tabular}{lcccccccc}
\toprule
   \multirow{2}{*}{\textbf{Encoder}} &\multirow{2}{*}{\textbf{Prompt}}& \multirow{2}{*}{\textbf{\makecell{Biasing \\ list size}}}  & \multicolumn{3}{c}{\textbf{test-clean}} & \multicolumn{3}{c}{\textbf{test-other}} \\
\cmidrule(lr){4-6} \cmidrule(lr){7-9}
& & & \textbf{WER} & \textbf{U-WER} & \textbf{B-WER} & \textbf{WER} & \textbf{U-WER} & \textbf{B-WER} \\
\midrule
  Pre-trained WavLM  &  \ding{55} & \ding{55} & 2.13 & 1.20 & 10.15 & 4.73 & 2.84 & 22.43 \\
  CTC Fine-tuned WavLM  & \ding{55} & \ding{55}  &2.11 & 1.20 & 10.02 & 4.20 & 2.43 & 20.76 \\
\midrule
\multirow{6}{*}{CTC Fine-tuned WavLM } & No bias &  \ding{55}&1.96 & 1.11 & 9.33 & 4.18 & 2.49 & 20.02 \\
\cmidrule(lr){2-9}
 & \multirow{4}{*}{Bias List} & 100&1.27 & 1.00 & 3.67 & 2.72 & 2.16 & 8.02 \\
 & &500 & 1.33 & 1.03 & 3.92 & 3.04 & 2.40 & 9.04 \\
 & &1000 & 1.33 & 1.00 & 4.16 & 2.99 & 2.31 & 9.33 \\
 & &2000 & 1.38 & 1.03 & 4.41 & 3.20& 2.47 & 10.02 \\
\cmidrule(lr){2-9}
& GT Hotwords &  \ding{55} & 1.13 & 0.94 & 2.78 & 2.68 & 2.32 & 6.00 \\
\bottomrule
\end{tabular}
}
\label{tab:wer}
\vspace{-3mm}
\end{table*}

\begin{table*}[h]
\centering
\caption{Performance comparison with different contextual ASR models using artificial biasing lists 
provided in~\cite{le2021contextualized} on Librispeech test sets. The biasing list size N is set to 1000.}
\begin{tabular}{lcccccc}
\toprule
\multirow{2}{*}{Model} & \multicolumn{3}{c}{test-clean} & \multicolumn{3}{c}{test-other} \\ \cmidrule(lr){2-4} \cmidrule(lr){5-7}
 & WER & U-WER & B-WER & WER & U-WER & B-WER \\ 
\midrule
CPPNet~\cite{huang2023contextualized} & 3.81 & 2.90 & 11.40 & 8.75 & 6.90 & 25.30 \\ 
Deep Biasing+BPB~\cite{sudo2024contextualized}& 3.47 & 3.00 & 7.70 & 7.34 & 6.40 & 15.80 \\ 
TCPGen+GNN enc. ~\cite{sun2022tree}& 3.10 & - & 6.70 & 7.90 & - & 17.80 \\ 
GA-CTC ~\cite{tang2024improving}& 2.40 & 2.00 & 6.30& 6.20 & 5.20 & 15.20 \\ 
TCPGen+p+phn-aware Q ~\cite{futami2024phoneme}& 2.20 & - & 4.60 & 6.00 & - & 12.30 \\ 
DB-NNLM ~\cite{le2021contextualized}& 2.14 & 1.60 & 6.70 & 6.35 & 5.10 & 17.20 \\ 
\textbf{Proposed Method}& \textbf{1.33} & \textbf{1.00} & \textbf{4.16 }& \textbf{2.99} & \textbf{2.31} &\textbf{9.33} \\ 
\bottomrule
\end{tabular}

\label{tab:compare}
\vspace{-4mm}
\end{table*}

\vspace{-2mm} 
\subsection{Filtering Algorithm}
Given a biasing list that includes named entities or rare words, such as a contact name list or a song playlist, we aim to employ a filtering mechanism to identify the most likely relevant hotwords with the sentence to be decoded. These selected hotwords are then incorporated into the prompt of the LLM to enhance speech recognition. 
We mainly utilize the coarse decoding results obtained from the WavLM Large model fine-tuned using CTC loss and decoded with the Viterbi algorithm, which we denote as ``inference sentence".

Firstly, we build a 2-gram index dictionary for the predefined biasing list. For example, assuming the biasing list is $[Bob, Joe]$, the 2-gram index dict will be $['bo':{'bob'}, 'ob':{'bob'}, 'jo':{'joe'}, 'oe':{'joe'}]$.  Next, we remove ``common words" from the inference sentence, which are defined as the top 5000 most frequently occurring words in the Librispeech training corpus. Then, we utilize the inference sentence to obtain preliminary screening candidates by indexing the dictionary. For instance, assuming the inference sentence is ``I like reading books", the candidate will be $['bob']$. Finally, for each rare word in the processed inference sentence, we calculate the edit distance to identify the most similar word from the candidate list and include it in the prompt.


\vspace{-2mm} 
\section{Experiments}

\vspace{-2mm} 
\subsection{Experimental Setup}
\subsubsection{Dataset}
We utilize the Librispeech corpus as our dataset, following previous literature~\cite{le2021contextualized,huang2023contextualized,tang2024improving,futami2024phoneme,sudo2024contextualized,sun2022tree}.
We finetune the official WavLM Large pre-trained model and train the LLM-based ASR model on the complete 960 hours of training data. The standard dev (dev-clean and dev-other) and test (test-clean and test-other) sets are adopted for performance evaluation. The artificial biasing list constructed in~\cite{le2021contextualized} is utilized for contextual ASR testing. They categorize the 5,000 most frequent words in the Librispeech training corpus as common words, with the remainder classified as rare words. The biasing list generated for the test set consists of two segments: rare words in the transcriptions, and distractors sampled from the 209.2K rare words vocabulary. Biasing lists of varying lengths are generated by incorporating $N=\{100, 500, 1000, 2000\}$ distractors into the lists. 

\vspace{-3mm} 
\subsubsection{Configuration}
\vspace{-2mm} 
\quad  \textbf{Fine-tuning WavLM Large:}
We fine-tune the pre-trained model for 80k steps on the 960 hours of training data. The learning rate increases linearly from 0 to 3e-5 in the initial 8k steps, stays constant for 32k steps, and then decays exponentially to 5\% of the peak rate in the remaining 40k steps. Parameters are frozen except for the final projection matrix for the initial 10k steps.
Model selection is based on the WER result on the dev-other subset. A batch contains at most 200 seconds of speech audio.

\textbf{LLM-based ASR Training and Inference:}
Following \cite{yang2024mala,ma2024embarrassingly}, the format of a training data input for the LLM is ``\textit{\textless speech\textgreater\ 
 USER: \textless prompt\textgreater\ ASSISTANT: \textless transcription\textgreater}''. \textit{\textless speech\textgreater} denotes the speech embedding, which has a dimension of 4096 aligned with the LLM. 
To steer the LLM towards performing ASR task, the \textit{\textless prompt\textgreater} is designed as ``\textit{Transcribe speech to text.}'' 
\textit{\textless transcription\textgreater} refers to the ground-truth transcription of the speech.

During the training process, we freeze the speech encoder (315.5M) and the LLM (6.7B) and only train the lightweight projector (15.7M). We compute the LM loss only on \textit{\textless transcription\textgreater}. The model is trained for 100k steps, with the learning rate increasing linearly from 0 to a peak of 1e-4 within the first 1k steps and then linearly decaying to 0 during the remaining training period. We utilize the AdamW optimizer with $\beta=(0.9,0.999)$ and zero weight decay. The experiments are conducted on 2 80GB A800 GPUs and the batch size is set to 4. 

During the inference process, the format of a test data input for the LLM is ``\textit{\textless speech\textgreater\ USER: \textless prompt\textgreater\ ASSISTANT: }''. Given this preceding input, the LLM will autoregressively generate the transcription. We utilize the beam search algorithm for decoding, with the beam size set to 4. 

\textbf{LLM-based Contextual ASR Training and Inference:}
The experimental setup is essentially consistent with LLM-based ASR, except that we need to modify the prompt to include a biasing list. Specifically, the \textit{\textless prompt\textgreater} template is modified to ``\textit{Transcribe speech to text. Some hotwords might help. The hotwords are \{\}}'', and will be further filled with the biasing words list.

In the training stage, we generate the biasing list dynamically from the transcriptions of the current training batch. Following \cite{pundak2018deep}, we set $P_{keep} = 0.5$ to enhance robustness for scenarios where no hotwords are provided. We set $N_{phrases} = 1$ and $N_{order} = 4$, resulting in biasing lists with an average length of 5 words. In the inference stage, we build the biasing list by filtering potentially relevant hotwords from the predefined hotwords list. Besides, we conduct experiments where we fill the prompt template with an empty string to validate the robustness of the proposed method in ``No bias" scenario. Additionally, we experiment with adding the ground truth hotwords to the prompt template as an upper bound for reference. The ground truth biasing list varies in length from 0 to over a dozen words, with an average of 2 words.


\vspace{-0.3cm}
\subsection{Experimental Results}

\begin{table*}[h!]
\centering
\caption{ Performance of our LLM-based contextual ASR model using different filtering methods with different biasing list size N on LibriSpeech test sets. Reported metrics are in the following format: WER (U-WER/B-WER). }
 \resizebox{\linewidth}{!}{
\begin{tabular}{|c|c|c|c|c|c|c|c|c|}
\hline
\multirow{2}{*}{\makecell{Filtering\\ Method}} & \multicolumn{2}{c|}{N=100} & \multicolumn{2}{c|}{N=500} & \multicolumn{2}{c|}{N=1000} & \multicolumn{2}{c|}{N=2000} \\ \cline{2-9} 
                       & test-clean & test-other & test-clean & test-other & test-clean & test-other & test-clean & test-other \\ \hline
F1           & \makecell{1.39 \\ (1.02 / 4.61)} & \makecell{3.19\\ (2.45 / 10.12)} & \makecell{1.48 \\ (1.06 / 5.05)} & \makecell{3.62 \\ (2.79 / 11.33)} & \makecell{1.54 \\ (1.10 / 5.32)} & \makecell{3.39 \\ (2.52 / 11.59)} & \makecell{1.64 \\ (1.16/ 5.79)} & \makecell{3.63 \\ (2.63 / 13.02)} \\ \hline
F2          & \makecell{1.40 \\ (1.00 / 4.85)} & \makecell{2.92\\ (2.18 / 9.78)} & \makecell{1.43 \\ (1.01 / 4.98)} & \makecell{3.39 \\ (2.60 / 10.76)} & \makecell{1.49\\ (1.05 / 5.28)} & \makecell{3.10 \\ (2.26 / 10.96)} & \makecell{1.53 \\ (1.05 / 5.67)} & \makecell{3.32\\ (2.40 / 11.88)} \\ \hline
F3          & \makecell{1.27 \\ (1.00 / 3.67)} & \makecell{2.72 \\ (2.16 / 8.02)} & \makecell{1.33 \\ (1.03 / 3.92)} & \makecell{3.04 \\ (2.40 / 9.04)} & \makecell{1.33 \\ (1.00/ 4.16)} & \makecell{2.99 \\ (2.31 / 9.33)} & \makecell{1.38 \\ (1.03 / 4.41)} & \makecell{3.20\\ (2.47 / 10.02)} \\ \hline

\end{tabular}
}
\label{tab:ablation}
\vspace{-3mm}
\end{table*}

\begin{table}[h!]
\vspace{-1mm}
\centering
\caption{Comparison of different filtering methods. }
\begin{tabular}{lccc}
\toprule
Filtering Method & F1 & F2 & F3 \\ 
\midrule
Similarity Threshold & \ding{51} & \ding{51} & \ding{55} \\ 
Remove Common Words & \ding{55} & \ding{51} & \ding{51} \\ 
Match for Each Word & \ding{55} & \ding{55} & \ding{51} \\ 
\bottomrule
\end{tabular}

\label{tab:filtering_methods}
\vspace{-3mm}
\end{table}

Tabel \ref{tab:wer} presents the performance of contextual speech recognition on the Librispeech dataset.
We employ WER, biased word error rate (B-WER), and unbiased word error rate (U-WER) as the evaluation metric, as the standard practice. B-WER is computed solely for words in the biasing list, while U-WER is calculated for words not in the biasing list.

For non-contextual ASR, our first model utilizes WavLM large pre-trained model as the encoder and gets WER of 2.13\%/4.73\% on the test-clean/test-other sets. When using the fine-tuned WavLM model as the encoder, the WER reduces slightly to 2.11\%/4.20\%, serving as the baseline for comparing the performance of contextual ASR. 

For contextual ASR, incorporating an empty string into the prompt during inference leads to WER/B-WER of 1.96\%/ 9.33\% and 4.18\%/20.02\% on the test-clean and test-other sets. The results are marginally better than the baseline, indicating the model's robustness and practicality for real-life scenarios where hotwords may not be available. 
Adding precise ground truth hotwords yields WER/B-WER of 1.13\%/2.78\% and 2.68\%/6.00\% on the test-clean and test-other sets. This represents an unattainable upper bound, serving as a reference value to measure the performance of our method.
In practical scenarios, a biasing list consisting of hundreds to thousands of words is provided, where only a few are beneficial ground truth hotwords that appear in transcriptions while the rest are obstructive distractor words. Experimental results indicate that the model performs best when the biasing list size is 100, which is intuitive. The WER/B-WER are 1.27\%/3.67\% for the test-clean set and 2.72\%/8.02\% for the test-other set, representing substantial relative reductions of 39.81\%/63.37\% and 35.24\%/61.37\% compared to the baseline, where U-WER remains largely unchanged. This clearly demonstrates the effectiveness of the filtering method and the model's efficient utilization of hotword information from the prompts. As the biasing list size increases, WER and B-WER show a slight rise. Nevertheless, even when the biasing list size grows to 2000, the WER/B-WER are 1.38\%/4.41\% on the test-clean set and 3.20\%/10.02\% on the test-other set, still representing distinct relative reductions of 34.60\%/55.99\% and 23.81\%/51.73\% compared to the baseline.

Table \ref{tab:compare} presents a performance comparison of different contextual ASR models using artificial biasing lists proposed in \cite{le2021contextualized} on Librispeech test sets, with the biasing list size set to 1000.

\vspace{-4mm} 
\subsection{Ablation Study}
\vspace{-2mm} 
The key to our LLM-based contextual ASR model is the filtering algorithm. The comparison of different filtering methods is illustrated in Table \ref{tab:filtering_methods}, and the evaluation results are presented in Table \ref{tab:ablation}.

All three filtering methods first involve building a 2-gram index dictionary for the predefined biasing list. For F1 filtering method, we use the inference sentence to retrieve candidate hotwords by indexing the dictionary. Next, we calculate the similarity score between each candidate and the inference sentence. Specifically, we compute the edit distance between the candidate and each word in the sentence and take the smallest value as the result; in other words, we calculate the ``shortest distance" between the candidate and the inference sentence. Finally, we rank the candidates based on their similarity scores in descending order. We then select either all words with a similarity score greater than 0.95 or the top five words (the average number of words in the biasing list during training).
Employing this filtering method, the WER/B-WER are relatively reduced by 34.12\%/53.99\% and 24.05\%/52.25\% on clean and test sets respectively compared to the baseline LLM-based ASR model when using a biasing list of 100. As the size of the biasing list grows, both WER and B-WER increase slightly but still show a noticeable reduction compared to the baseline.

The experimental results reveal an issue with this filtering approach. Due to the unknown specific locations of rare words within the inference sentence, we need to compute the similarity between the candidate and each word of the sentence, leading to unnecessary matching of bias words highly similar to common words in the sentence. Thus, for F2 filtering method, we utilize the dataset's provided 5000 common words list to remove common words from the inference sentence, before retrieving initial candidate hotwords from the index dictionary. This improvement results in a relative reduction of 7.82\%/6.41\% in WER/B-WER on the test-other set, and a modest reduction of 4.44\% and 1.40\% on the test-clean set, except for a minor performance drop for biasing list of 100.

The limitation of F2 filtering method is that, while the inference sentence only contains a few rare words, solely using a similarity threshold for filtering may result in the highest-scoring hotwords being similar to some of these rare words, leaving other rare words unmatched with any relevant hotwords. To address this, the F3 filtering method matches every word in the inference sentence with the most similar hotword from the biasing list. This refinement shows a significant enhancement over the F2 filtering method. On the test-clean set, WER/B-WER is relatively reduced by 9.21\%/22.26\%, while on the test-other set, WER/B-WER is relatively reduced by 6.08\%/16.13\%. Ultimately, we adopt the F3 filtering method as our proposed approach.

\vspace{-4mm} 
\section{Conclusion}
\vspace{-3mm} 

In this work, we extend the LLM-based ASR model to contextual ASR tasks. We introduce an effective filtering algorithm designed to retrieve truly relevant hotwords. Our experiment results demonstrate that the CTC-Assisted LLM-Based Contextual ASR model significantly outperforms the baseline, improving the performance significantly compared to the previous methods. This underscores the accuracy and effectiveness of our filtering method, further validating that large language models can efficiently incorporate hotwords from prompts to enhance speech recognition accuracy, serving as a highly effective framework for performing contextual ASR tasks.
In the future, we will further explore LLM-based contextual ASR and attempt to adapt and optimize our method to other datasets, and different languages other than English.

\section{Acknowledgements}
\vspace{-3mm} 
This work was supported by the National Natural Science Foundation of China  (No. 62206171 and No. U23B2018), Shanghai Municipal Science and Technology Major Project under Grant 2021SHZDZX0102, the International Cooperation Project of PCL.

\vspace{-4mm} 
\bibliographystyle{IEEEbib}
\bibliography{refs}
\end{document}